\documentclass[11pt]{article}
\newsavebox{\PSLASH}
\sbox{\PSLASH}{$p$\hspace{-1.8mm}/}

\input{amssym}
\begin{document}
\title{On the thin wall limit of thick planar domain walls}
\author{S. Ghassemi$^{1}$\footnote{e-mail: sima\_gh@mehr.sharif.edu},
S. Khakshournia$^{2}$\footnote{e-mail: skhakshour@aeoi.org.ir},
and R. Mansouri$^{3}$\footnote{e-mail: mansouri@sharif.edu} \\ \\
$^{1,3}$Department of Physics, Sharif University of Technology,\\
Tehran
11365-9161, Iran\\
$^{2}$Nuclear Research Center, Atomic Energy Organization of Iran,
 Tehran, Iran}\maketitle
 \[\]
 \[ \]
\begin{abstract}
Considering a planar gravitating thick domain wall of the $\lambda
\phi^4$ theory, we demonstrate how the Darmois junction conditions
written on the boundaries of the thick wall with the embedding
spacetimes reproduce the Israel junction condition across the wall
when one takes its thin wall limit.

\end{abstract}
\newpage
\section{Introduction}
Domain walls are solutions to the coupled Einstein-scalar field
equations with a potential having a spontaneously broken discrete
symmetry and a discrete set of degenerate minima. In the simplest
case of two minima, a domain wall having a non-vanishing energy
density appears in the separation layer, with the scalar field
interpolating between these two values.\\
Domain walls in the cosmological context have a long history
\cite{vilen}. It was realized very early that the formation of
domain walls with a typical energy scale of $\geq 1 MeV$ must be
ruled out \cite{zel}, because a network of such objects would
dominate the energy of the universe, violating the observed isotropy
and homogeneity. Domain walls were reconsidered in a possible late
time phase transition scenario at the scale of $\leq 100 MeV$. Such
walls were supposed to be thick because of the low temperature of
the phase transition \cite{hill}. The suggestion that Planck size
topological defects could be regarded as triggers of inflation,
revived the discussion of thin and thick domain walls \cite{vil2}.
The realization of our universe as a $(3+1)$-dimensional domain wall
immersed in a higher dimensional spacetime
has led to the recent numerous studies \cite{rand}.\\
The first attempts to investigate the gravitational properties of
domain walls were based on the so called thin wall approximation. In
this approach one forgets about the underlying field theory and
simply treats the domain wall as a zero thickness (2+1)-dimensional
timelike hypersurface embedded in a four-dimensional spacetime. The
Israel thin wall formalism \cite{israel} is then used to continue
the solutions of the Einstein equations on both sides of the wall in
the embedding spacetime across the thin wall. However, such
spacetimes have delta function-like distributional curvature and
energy-momentum tensor supported on the hypersurface. Using this
procedure, the first vacuum solutions for a spacetime containing a
infinitely thin planar domain wall was found by Vilenkin
\cite{vil3}, and Ipser and Sikivie \cite{ipser}. The very
interesting feature of such domain walls is that they are not
static, but have a de Sitter-like expansion in the wall's plane.
External observers experience a repulsion from the wall, and there
is an event horizon at finite proper distance from the wall's core.
These results were initially obtained within the framework of the
Israel thin wall formalism which has been shown to be an
approximative description of a real thick wall by using an expansion
scheme in powers of the wall thickness \cite{widrow,garfin}.
Typically, a self gravitating domain wall has two length scale, its
thickness $w$ and the distance to event horizon which can be
compared to $w$. Since these lengths are expressed in terms of the
coupling constants of the theory, thin walls turn out to be an
artificial construction in terms of these
underlying parameters as mentioned in \cite{gre99}.\\
The first exact dynamical solution to thick planar domain walls
was obtained by Goetz \cite{goetz} and later a static solution was
recovered with the price of sacrificing reflection symmetry
\cite{mukh}. Within the context of a fully nonlinear treatment of
a scalar field coupled to gravity, Bonjour, Charmousis and Gregory
(BCG) found an approximate but analytic description of the
spacetime of a thick planar domain wall of the $\lambda \phi^4$
model by examining the field equations perturbed in a parameter
characterizing the gravitation interaction of the scalar field
\cite{gre99}. Recently, the thin wall limit of Goetz's solution
was studied in \cite{rom} and it has been shown that this solution
has a well-defined limit. But to date the thin wall
limit of BCG's solution has never been investigated.\\
In this paper, we study the thin wall limit of the thick planar
domain wall described by BCG spacetime. To do so, we use the
formalism developed by Mansouri and Khakshournia (MK)
\cite{khak02} to treat dust thick shells.\\
The organization of this paper is as follows. In section 2 we give a
brief introduction to BCG thick wall solution and summarize all the
useful equations we will need in the present work . In Section 3 we
describe the thick wall formalism of reference \cite{khak02} and
apply it to the thick planar domain wall solved by BCG. Section 4
considers the thin wall limit of the thick domain wall solution
followed by the conclusion.

\section{The thick planar domain wall solution}
Domain walls as the regions of varying scalar field separating two
vacua with different values of field are usually described by the
matter Lagrangian:
\begin{equation}\label{Lagrangian}
L=\nabla_{\mu}\phi\nabla^{\mu}\phi-V(\phi),
\end{equation}
where $\phi$ is a real scalar field and $V(\phi)$ is a symmetry
breaking potential which we take to be
$V(\phi)=\lambda(\phi^2-\eta^2)^2$, where $\lambda$ is a coupling
constant and $\eta$ the symmetry breaking scale. Looking for a
static solution of the equation of motion derived from this
lagrangian in flat space-time, we get
\begin{equation}\label{phi}
X=\tanh(\frac z w),
\end{equation}
where $X=\frac\phi\eta$, $w=\frac{1}{\sqrt{\lambda}\eta}$, and z is
the coordinate normal to the wall. This particular solution
represents an infinite planar domain wall centered at $z=0$. From
the stress-energy tensor of the wall, one can easily observe that
the wall energy density peaked around $z=0$ falls down effectively
at $z=w$. So $w$ a length scale in the system is called the
effective
thickness of the wall within the theory.\\
We now look at the planar gravitating domain wall solutions. The
line element of a plane symmetric spacetime may be written in the
general form
\begin{equation}\label{metric}
ds^2=A^2(z)dt^2-B^2(z,t)(dx^2+dy^2)-dz^2,
\end{equation}
which displays reflection symmetry around the wall's core located at
z=0, where z is the proper length along the geodesics orthogonal to
the wall. In order to obtain a thick domain wall solution one should
solve the coupled system of the Einstein and scalar matter field
equations as follows
\begin{equation}\label{Richi}
R_{\mu\nu}=8\pi
G\eta^2\left(2X_{,\mu}X_{,\nu}-\frac{1}{w^2}g_{\mu\nu}(X^2-1)^2\right),
\end{equation}
\begin{equation}\label{field}
\square X+\frac{2}{w^2}X(X^2-1)=0,
\end{equation}
where $R_{\mu\nu}$ is the spacetime Ricci tensor. For a static
field $X(z)$, Einstein equations (\ref{Richi})
constraint $B(z,t)$ as $B(z,t)=A(z)\exp(kt)$.\\
BCG investigated the spacetime of a thick gravitating planar domain
wall for a $\lambda \phi^4$ potential \cite{gre99}. In the context
of their work a dimensionless parameter $\epsilon$ arisen from
equation (\ref{Richi}) is singled out to characterize the coupling
of gravity to the the scalar field namely
\begin{equation}\label{epsilon}
\epsilon=8\pi G\eta^2.
\end{equation}
Supposing that gravity is weakly coupled to the scalar field,
$A_0(z)$ and $X(z)$ may be expanded in the powers of $\epsilon$:
\begin{eqnarray}
A(z)=A_0(z)+\epsilon A_1(z)+O(\epsilon^2),\\\
X(z)=X_0(z)+\epsilon X_1(z)+O(\epsilon^2).
\end{eqnarray}
In the $\epsilon\rightarrow  0$ limit, the results should be the
same as non-gravitating planar wall's which are  $A(z)=1$ and
$X(z)=\tanh(\frac z w)$. Using these expansions, BCG solved the
coupled Einstein and scalar matter field equations to first order
in $\epsilon$ and obtained the following results:
\begin{equation}\label{inside metric}
A_i(z)=1-\frac{\epsilon}{3}\left[2\ln[\cosh(\frac z w)]+\frac 1 2
\tanh^2(\frac z w)\right]+O(\epsilon^2),
\end{equation}
\begin{equation}\label{inside metric2}
k_i=\frac 2 3\frac{\epsilon}{w}+O(\epsilon^2),
\end{equation}
\begin{equation}\label{inside metric3}
X_i(z)=\tanh(\frac z w)-\frac{\epsilon}{2}{\rm sech}^2(\frac z w
)\left[\frac z w+\frac 1 3 \tanh(\frac z w )\right]+ O(\epsilon^2).
\end{equation}
Thus, (\ref{inside metric}) is a perturbative solution to the
spacetime of the thick wall (\ref{metric}) obtained by BCG. In the
following sections we will use this solution.
\section{The thick wall formalism}
In this section we first make a short review of the thick wall
formalism developed by  MK in \cite{khak02}. Then we apply it to the
thick planar domain wall described by the metric (\ref{metric}) . In
MK formalism a thick wall is modelled with two boundaries $\Sigma_1$
and $\Sigma_2$ dividing a spacetime $\cal M$ into three regions. Two
regions $\cal M_{+}$ and $\cal M_{-}$ on either side of the wall and
region $\cal M$$_{0}$ within the wall itself. Treating the two
surface boundaries $\Sigma_1$ and $\Sigma_2$ separating the manifold
$\cal M$$_{0}$ from two distinct manifolds $\cal M_{+}$ and $\cal
M_{-}$, respectively, as nonsingular timelike hypersurfaces, we do
expect the intrinsic metric $h_{\mu\nu}$ and extrinsic curvature
tensor $K_{\mu\nu}$ of $\Sigma_j\hspace{0.2cm}$ (j=1,2) to be
continuous across the corresponding hypersurfaces. These
requirements named the Darmois conditions are formulated as
\begin{equation}\label{hmn}
[h_{\mu\nu}]_{\Sigma_j}=0\hspace {1cm} j=1,2,
\end{equation}
\begin{equation}\label{thick israel}
[K_{\mu\nu}]_{\Sigma_j}=0\hspace {1cm} j=1,2,
\end{equation}
where the square bracket denotes the jump of any quantity that is
discontinuous across $\Sigma_j$. To apply the Darmois conditions on
two surface boundaries of a given thick wall one needs to know the
metrics in three distinct spacetimes $\cal M_{+}$, $\cal M_{-}$ and
$\cal M$$_{0}$ being jointed at $\Sigma_j$. While the metrics in
$\cal M_{+}$ and $\cal M_{-}$ are usually given in advance, knowing
the metric in the wall spacetime $\cal M$$_{0}$ requires
a nontrivial work. \\
Let us now impose these junction conditions for a self gravitating
thick planar domain wall described in the previous section.
Recalling $w$ is the effective thickness of the wall, we first
follow Ref. \cite{garfin} to introduce a parameter $\Delta \gg1$ to
assure that the scalar field takes its vacuum values on the wall
boundaries  $\Sigma_1$ and $\Sigma_2$ being located at the proper
distances $z=\pm \:\: \Delta w/2$ far from the wall's core surface
at $z=0$. We can then think of $\Delta w$ as the proper thickness of
the planar domain wall. In the coordinate frame of the metric
(\ref{metric}) in which the wall is stationary, the nonvanishing
components of the intrinsic metric $h_{\mu\nu}$ and extrinsic
curvature $K_{\mu\nu}$ of $\Sigma_j$ take the following simple
forms:
\begin{eqnarray}\label{hcompon}
h_{\mu\nu}=g_{\mu\nu},\hspace {1cm}  \mu,\nu\neq z,
\end{eqnarray}
\begin{equation}\label{Kcompon}
K_{\mu\nu}=-\frac 1 2 g_{\mu\nu,z}.
\end{equation}
In order to find the spacetime metric on both sides of $\Sigma_j$,
we first note that within the vacuum region $\cal M_{+}$ ($\cal
M_{-}$) in which $\phi= \eta(-\eta)$, the spacetime metric can be
easily determined by solving the Einstein equations (\ref{Richi})
yielding
\begin{equation}\label{vacuum metric}
A_o(z)=-k_o |z|+ C.
\end{equation}
Using (\ref{hcompon}) and (\ref{Kcompon}) we write down junction
conditions (\ref{hmn}) and (\ref{thick israel}) as
\begin{equation}\label{k}
k_i=k_0,
\end{equation}
\begin{equation}\label{metric constraints1}
A_i(z)|_{z=\Delta w/2}=A_o(z)|_{z=\Delta w/2},
\end{equation}
\begin{equation}\label{metric constraints2}
\frac{\partial A_i(z)}{\partial z}|_{z=\Delta w/2}=\frac{\partial
A_o(z)}{\partial z}|_{z=\Delta w/2}.
\end{equation}
We now use the solutions (\ref{inside metric}) and (\ref{vacuum
metric}) due to BCG for the wall metric in the region $\cal
M$$_{0}$, where the scalar field varies according to (\ref{inside
metric3}), and for the metric in the vacuum regions, respectively.
Then the junction conditions (\ref{metric constraints1}) and
(\ref{metric constraints2}) lead to the following constraints on the
vacuum metric constants $C$ and $k_o$
\begin{equation}\label{c}
C=1+k_o\frac{\Delta w
}{2}-\frac{\epsilon}{3}\left(2\ln(\cosh(\Delta/2))+\frac 1 2
\tanh^2(\Delta/2)\right),
\end{equation}
\begin{equation}\label{ko}
k_o=\frac {\epsilon}{w}\left(\tanh(\frac{\Delta}{2})-\frac 1 3
\tanh^3(\frac{\Delta}{2})\right).
\end{equation}
Note that within the context of BCG work, it is supposed that the
boundaries of the wall where the scalar field takes its vacuum
values are at infinity. But here we have modelled the thick planar
wall in such a way that the wall boundaries $\Sigma_j$ are situated
at the finite proper distances $\pm\Delta\,w/2$ from the core of the
wall. Hence, we choose $\Delta$ to be sufficiently large in order to
simulate BCG solution within our wall model.
\section{From the thick to thin domain walls}
We now turn our attention to the thin wall limit of our thick wall
model. First let us define the process of passing from a thick
gravitating domain wall to a thin one by letting $\epsilon$ and $w$
go to zero while keeping their ratio $\frac {\epsilon}{w}$ fixed.
This has the effect that the distance of the event horizon to the
domain wall remains finite.\\ We then rewrite the Darmois junction
condition (\ref{thick israel}) demanding the continuity of the
extrinsic curvature tensor $K_{\mu\nu}$ across the thick wall
boundary, say $\Sigma_1$, located at the proper distance $z=\Delta
w/2$ as
\begin{equation}\label{kmunu}
K^o_{\mu\nu}|_{z=\Delta w/2}=K^i_{\mu\nu}|_{z=\Delta w/2}.
\end{equation}
From the formula (\ref{Kcompon}) one can evaluate the right hand
side of the $(tt)$ component of the equation (\ref{kmunu}) for the
metric (\ref{metric}) using the BCG wall metric solution
 (\ref{inside metric}). This yields
\begin{equation}\label{ktt}
K^o_{tt}|_{z=\Delta
w/2}=\frac{\epsilon}{w}\left(\tanh(\frac{\Delta}{2})-\frac 1 3
\tanh^3(\frac{\Delta}{2})\right)\left(1-\frac{\epsilon}{3}\left[(2\ln[\cosh(\frac
\Delta 2)]+\frac 1 2 \tanh^2(\frac \Delta 2)\right]\right).
\end{equation}
Imposing the above thin wall limit prescription, the equation
(\ref{ktt}) reduces to
\begin{equation}\label{ktt2}
K^o_{tt}|_{z=0}=\frac
{\epsilon}{w}\left(\tanh(\frac{\Delta}{2})-\frac 1 3
\tanh^3(\frac{\Delta}{2})\right).
\end{equation}
To identify the right hand side of the equation (\ref{ktt2}) we
recall the definition of the surface energy density $\sigma$ of an
infinitely thin wall. Within our thick wall model it takes the
form
\begin{eqnarray}\label{sigmap}
\sigma=\lim_{(w\rightarrow 0, \epsilon \rightarrow 0)}\int^{\Delta
w/2}_{-\Delta w/2}\rho dz,
\end{eqnarray}
where $\rho=\rho(z)$ is the energy density of the scalar field
which is computed for the BCG scalar field solution (\ref{inside
metric3}). Finally, we get the following expression for $\sigma$
\begin{eqnarray}\label{sigmap2}
\sigma=\frac{\epsilon}{2\pi G w}\left(\tanh(\frac{\Delta}{2})-\frac
1 3 \tanh^3(\frac{\Delta}{2})\right),
\end{eqnarray}
where we used the definition $\epsilon$ given by (\ref{epsilon}).
Comparing the results (\ref{ktt2}) and (\ref{sigmap2}) one
immediately obtains
\begin{eqnarray}\label{thin}
K^o_{tt}|_{z=0}=2\pi G\sigma.
\end{eqnarray}
Now, consider a planer thin domain wall placed at $z=0$. The Israel
thin wall approximation treats the wall as a singular hypersurface
with the surface energy $\sigma$ separating the two plane symmetric
vacuum spacetimes $\cal M_{+}$ and $\cal M_{-}$ from each other.
Then the Israel junction condition across the wall is written as
\begin{equation}\label{israelp}
K^o_{\mu\nu}=2\pi G\sigma h_{\mu\nu}|_{z=0}.
\end{equation}
Not surprisingly, we now see that the equation (\ref{thin}) is just
the same as the $(tt)$ component of the Israel's equation
(\ref{israelp}), since from the equations (\ref{hcompon}) and
(\ref{vacuum metric}) it follows that $h_{tt}|_{z=0}=A_o^2=C$, where
in the thin wall limit, the equation (\ref{c}) reduces to $C=1$.
Using the condition (\ref{k}) being held at $z=0$, one can easily
show the same results for the $(xx)$ and $(yy)$ components of the
equation (\ref{kmunu}).
\section{Conclusion}
We have studied the thin wall limit of the thick planar domain
wall solution obtained by Bonjour, Charmousis and Gregory in Ref.
\cite{gre99}. Treating the thick planar wall as a defect having
two boundaries at the same proper distance from the wall's core,
as formulated by Mansouri-Khakshournia for the case of a dust
shell in Ref. \cite{khak02}, we have shown that the Darmois
junction conditions for the extrinsic curvature tensor at the wall
boundaries with the two embedding spacetimes generate the
well-known Israel jump condition at the separating boundary of the
corresponding thin wall with the same embedding spacetimes. We
have realized that in the process of passing from a thick  planar
domain wall to the thin one all the information about the internal
structure of the wall are squeezed  in the parameter $\sigma$
characterizing the wall surface energy density as introduced
in (\ref{sigmap2}).\\\\
{\large \bf Acknowledgement}: We would like to thank S. Khosravi
for useful discussions.


\begin{thebibliography}{99}
\bibitem{vilen}
A. Vilenkin, Phys.Rev. D23, 852 (1981).
\bibitem{zel}
Ya, B. Zel'dovich,I. Yu. Kobzarev, and L. N. okun, Sov. Phys. JETP
40,1 (1974).
\bibitem{hill}
C. T. Hill, D. N. Schramm, J. N. Fray, Comm. Nucl. Part. Sci. 19, 25
(1989).
\bibitem{vil2}
A. Vilenkin, Phys. Rev. Lett. 72, 3137 (1994).

\bibitem{rand}
L. Randall, R Sundrum, Phys. Rev. Lett. 83, 3370 (1999); 83, 4690
(1999).
\bibitem{israel}
W. Israel, Nuovo Cimento B 44, 1 (1966).
\bibitem{vil3}
A. Vilenkin, Phys. Lett. 133B, 177 (1983).
\bibitem{ipser}
J. Ipser and P. Sikivie, Phys. Rev. D 30, 712 (1984).
\bibitem{widrow}
L. M. Widrow, Phys. Rev. D 39, 3571 (1989); 39, 3576 (1989).
\bibitem{garfin}
D. Garfinkle and R. Gregory, Phys. Rev. D 41, 1989 (1990).
\bibitem{gre99}
F. Bonjour, C. Charmousis and R. Gregory, Class. Quantum Grav. 16,
2427 (1999),\\
\hspace{1.8cm}[gr-qc/9902081].
\bibitem{goetz}
G. Goetz, J. Math. Phys. 31, 2683 (1990).
\bibitem{mukh}
M. Mukherjee, Class. Quantum Grav. 10, 131 (1993). \textbf{16},
2427 (1999).
\bibitem{rom}
R. Guerrero, A. Melfo and N. Pantoja, Phys. Rev. D 65, 125010
(2002), [gr-qc/0202011].
\bibitem{khak02}
S. Khakshournia and R. Mansouri, Gen. Rel. Grav. \textbf{34}, 1847
(2002), [gr-qc/0308025]

\end{thebibliography}
\end{document}